# Towards a Framework for Openness in Foundation Models

***Proceedings from the Columbia Convening on Openness in Artificial Intelligence***

Adrien Basdevant,[1, 13] Camille François,[2] Victor Storchan,[3] Kevin Bankston,[4] Ayah Bdeir,[1] Brian Behlendorf,[5] Merouane Debbah,[6] Sayash Kapoor,[7] Yann LeCun,[8] Mark Surman,[1] Helen King-Turvey,[14] Nathan Lambert,[10] Stefano Maffulli,[11] Nik Marda,[1] Govind Shivkumar,[12] Justine Tunney[1,14].

[1]Mozilla, [2]Columbia University, [3]Mozilla.ai, [4]Center for Democracy and Technology, [5]Linux Foundation, [6]Technology Innovation Institute, [7]Princeton University, [8]Meta, [9]Philanthropy Matters, [10]Allen Institute for AI, [11]Open Source Initiative, [12]Omidyar Network, [13]Entropy Law, [14]Llamafile.
Correspondence to Camille Francois (cmf2157@columbia.edu) and Ayah Bdeir (ayah@mozillafoundation.org)

## Abstract

Over the past year, there has been a robust debate about the benefits and risks of "open sourcing" foundation models. However, this discussion has often taken place at a high level of generality or with a narrow focus on specific technical attributes. In part, this is because defining "open source" for foundation models has proven tricky, given its significant differences from traditional software development. In order to inform more practical and nuanced decisions about opening AI systems, including foundation models, this paper presents a framework for grappling with openness across the AI stack. It summarizes previous work on this topic, analyzes the various potential reasons to pursue openness, and outlines how openness varies in different parts of the AI stack, both at the model and at the system level. In doing so, we hope to provide a common descriptive framework to deepen a nuanced and rigorous understanding of openness in AI and enable further work around definitions of openness and safety in AI.

## I. Introduction and Context

Openness in general and open source in particular have played a core role[1] in advancing innovative software development, including in artificial intelligence, and openness must continue to be defended as a global priority.[2] Some have argued that openness in foundation models (FMs) is a primary driver for societal threats, rather than an opportunity for advancing research and strengthening safety, and have thus asserted that a key lever for tackling risks from FMs — such as disinformation and non-consensual intimate imagery — is by restricting access to open models.[3] On the other hand, other scholars have contended that a wider range of governance tools would be more effective in mitigating those marginal risks in practice, while also preserving the myriad of benefits that openness in FMs can provide to society.[4] As the EU AI Act and U.S. reporting requirements on FMs — also referred to as general-purpose AI models (GPAIMs) — move toward implementation, there is an increasingly urgent need for more nuanced and informed explorations.

[1] Hoffmann, Nagle, Zhou, "The Value of Open Source Software." *Harvard Business School Strategy Unit Working Paper*, January 2024.
[2] See, e.g., Mozilla, "Joint Statement on AI Safety and Openness," October 2023.
[3] See, e.g., Seger, Dreksler, Moulange, Dardaman, Schuett, Wei, et al, "Open-Sourcing Highly Capable Foundation Models: An Evaluation of Risks, Benefits, and Alternative Methods for Pursuing Open-Source Objectives", Centre for the Governance of AI, September 2023.

It is challenging to define openness in the context of FMs, as definitions of "open source" software do not translate easily to AI systems. However, this work is vital for developing shared understandings, norms, benchmarks, and best practices for organizations developing FMs, and policymakers and civil society working to unlock the benefits and mitigate the risks from FMs. This work is also necessary for understanding the range of potential design choices around openness across the AI stack — and ensuring that this conversation goes beyond a narrow focus on model weights.

This paper focuses on surveying existing approaches to defining openness in AI models and systems. It proposes a descriptive framework to facilitate an analysis of how each component across the FM stack contributes to openness, and to enable normative definitions of openness in AI. It intentionally does not present a definitive list of requirements for openness.

In February 2024, Mozilla and the Columbia Institute of Global Politics brought together over 40 leading scholars and practitioners working on openness and AI for the Columbia Convening[5]. These individuals — spanning prominent open source AI startups and companies, non-profit AI labs, and civil society organizations — focused on exploring what "open" should mean in the AI era. After a subgroup developed a background document to help alignment and analysis, these individuals met in person for a workshop focused on developing a framework for openness in AI, and this paper reflects many of their insights and key takeaways. We also intend for this paper to enable further work on this topic, including work to develop stronger safety safeguards for open systems.

[4] See, e.g., Bommasani, Kapoor, et al., "Considerations for Governing Open Foundation Models," Stanford Institute for Human-Centered AI, December 2023. For additional perspectives on this debate, see the February 2024 solicitation for comments from the U.S. Department of Commerce National Telecommunications and Information Administration, including submissions from the Center for Democracy and Technology, HuggingFace, Meta, and Mozilla which all reference the work of the Columbia Convening on Openness and AI.

[5] For more information about this convening, and for a full list of participants, see Ayah Bdeir and Camille François, "Introducing the Columbia Convening on Openness and AI," Mozilla.

## 2. Motivations

Broadly, there are a handful of timely areas of exploration on openness and AI, notably: 1. reflecting on the impact of **open source on AI development** and 2. understanding the intricacies of **opening up AI itself**, including determining which aspects of the technological stack *could* be made accessible and in what manner. This work focuses on the latter.

The literature provides diverging perspectives on what openness can and should mean in the context of AI. The authors of this paper argue that the following four tenets provide a robust foundation for a nuanced exploration of openness and AI:

1. Openness must be considered both at the model stack and the system stack levels. Examinations that are either overly focused on model weights or release strategies may overlook the broader considerations that shape an AI system's benefits and risks in practice.
2. There will be different "forms" of openness in AI systems. This leaves room for a granularity of terminology (e.g., open science, open data, open weights, available weights), and new normative criteria (e.g. in Fig. 12, appendix III, we illustrate how the Model Openness Framework can be viewed as a normative layer on the descriptive framework we propose in this paper).
3. There is a timely need to better agree the benefits, potential risks and modalities of "opening up" the different components and attributes of the stack across AI systems.
4. Safety must be a core consideration. Safety cannot be thoroughly addressed by merely scrutinizing AI models. The context and deployment environment are also essential to ensuring AI is safe, making it vital to differentiate between open models and open AI systems. The latter can incorporate additional safeguards and introduce additional risks that may not be present in AI models alone.

With these fundamentals in mind, this paper proposes **a nuanced framework** (page 4, **Fig. 1**) that identifies "dimensions of openness", with an emphasis on AI systems with model weights that are downloadable or more open. This paper also invites a discussion on the key tradeoffs for leveraging openness and how the different dimensions of openness relate to the different goals one may want to pursue (see **Fig. 6**). This framework does not purport to provide a complete exposition of openness in AI, but rather, is oriented toward being practical for developers, policymakers, and other key stakeholders.

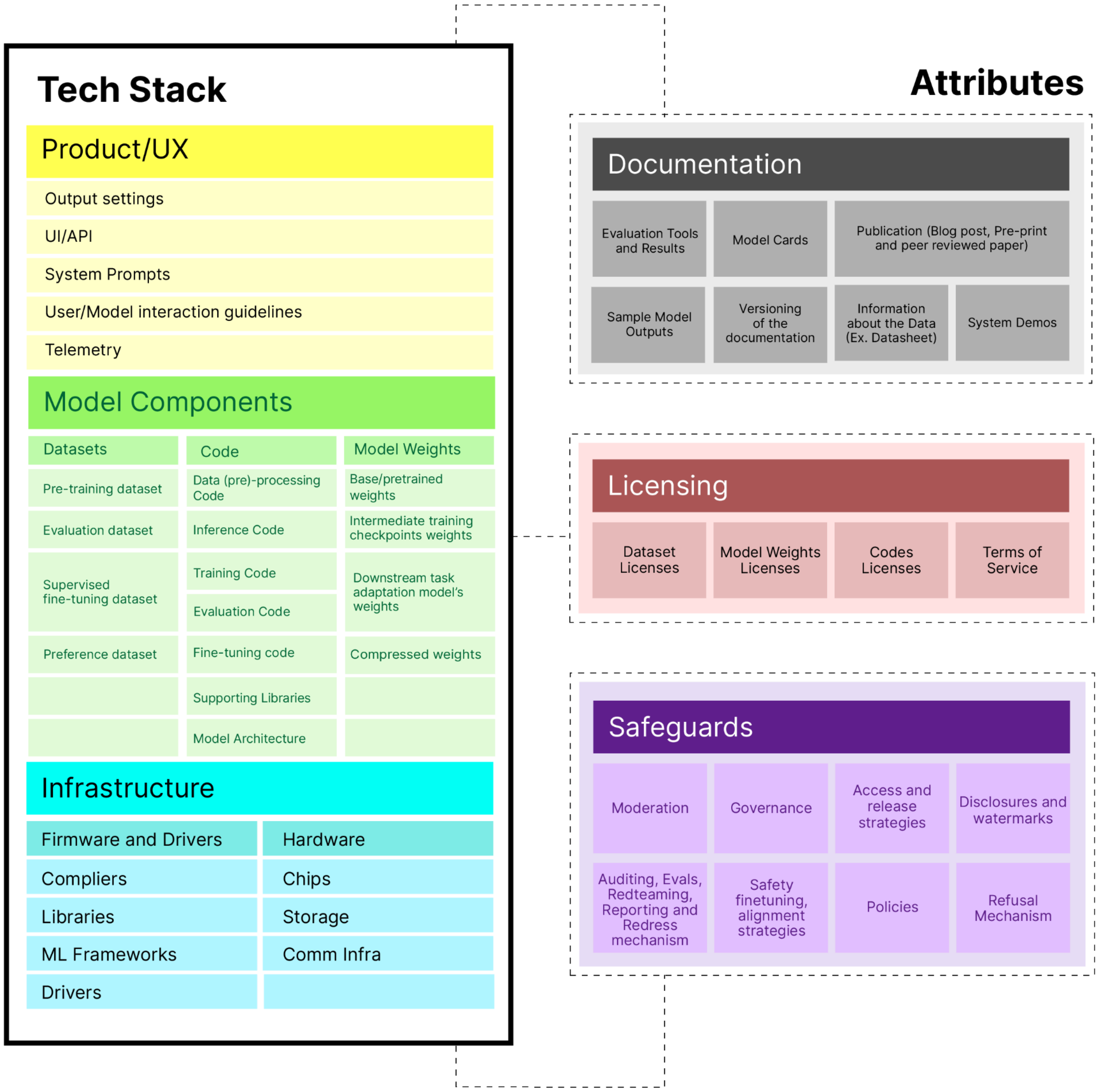

**Fig. 1 : The Framework - General-purpose AI system stack & Dimensions of Openness**

Prior relevant research was also surfaced and collectively analyzed throughout the Columbia Convening to help participants reach productive alignment on the difficulties and promises of defining openness in AI systems. This covers the diversity of underlying motivations for pursuing AI openness and previous work aiming to define openness for AI — which we organize as approaches that define openness as a gradient ("more or less open"), as a score (e.g., "7 out of 10 in openness"), or as a binary (either "open" or "closed"). Finally, this paper includes three appendices. **Appendix I** highlights the different parts of the foundation model stack, as well as corresponding attributes and the potential benefits of openness in each component. **Appendix II** highlights ongoing questions and areas of disagreement. And, **Appendix III** engages with and integrates work of the Model Openness Framework led by the Linux Foundation.

## 3. The Framework

This section outlines a framework focused on AI systems with model weights that are at least downloadable. It begins with a representation of an AI model stack (3.1.), and then zooms out to an AI system stack (3.2.) to common attributes applicable to the whole stack (3.3.) to the final framework (3.4.) so that it can be read and used in the light of the different benefits of AI openness (3.5.). Illustrations of a more in depth review of opening the foundation model stack is in Appendix I

### 3.1. AI Model Stack

Reaching an agreement on a definition for AI has already proven to be complicated.[6] Understanding openness in AI is a complex task as well. Indeed, AI models are not just code; they are trained on massive datasets, deployed on intricate computing infrastructure, and accessed through diverse interfaces and modalities.

To effectively depict an AI system stack, it may be beneficial to initially focus on the AI model tech stack. To begin with, the figure below presents the core components that are particularly relevant for considerations around openness of a general-purpose AI model (GPAIM, **Figure 2**), to then reflect on a general-purpose AI system (GPAIS, Figure 3).

At the model level, there are three main artifacts: Datasets, Code, and Model weights. Each of these artifacts has sub-components. It is important to note that all sub-component are not mandatory/applicable in every configuration (e.g. some models may have reward model's weights or preference dataset, others do not). Furthermore, the taxonomy of sub-components may vary based on the specific model type. Given this variability, it becomes crucial to tailor the AI model stack to accommodate different modalities, such as input and output types.[7]

The below Figure illustrates these various layers at the model level.

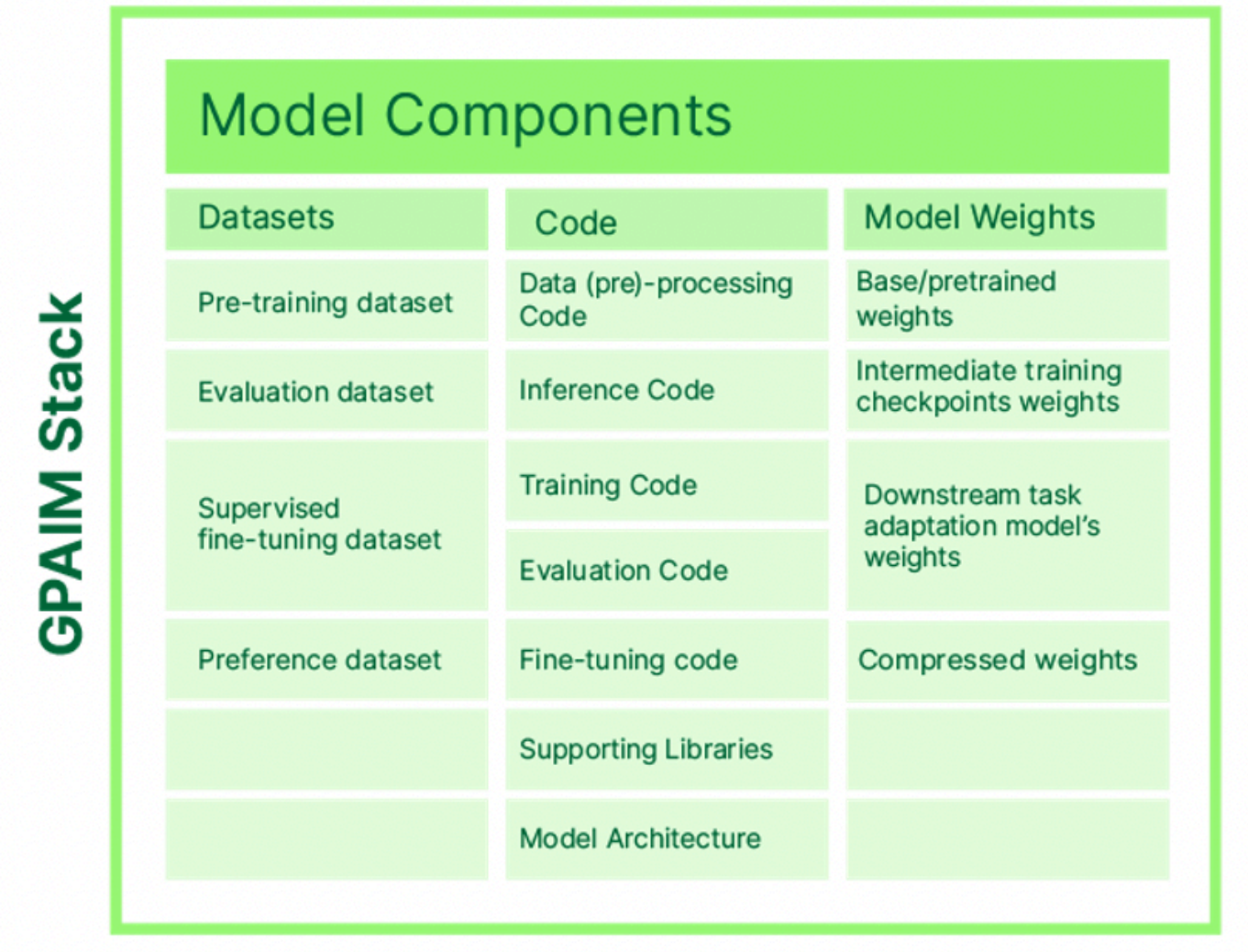

**Fig. 2 : General Purpose AI Models (GPAIM) Tech Stack**

### 3.2. AI System Stack

It is critical to distinguish between open models, on the one hand, and open AI systems or products, on the other hand.

The EU AI Act decided not to use the term "foundation model" and instead used "general-purpose AI model" (with "large generative AI model" being one example of them)[8]. More precisely, it introduces an important nuance in between “general-purpose AI model” (GPAIM) and “general-purpose AI system” (GPAIS), stating that “*Although AI models are essential components of AI systems, they do not constitute AI systems on their own. AI models require the addition of further components, such as for example a user interface, to become AI systems*” (Recital 60a). This distinction is taken into consideration.

[6] See, e.g., Stuart Russell, Karine Perset, Marko Grobelnik, "Updates to the OECD's definition of an AI system explained," Organisation for Economic Co-operation and Development, Nov. 2023.

[7] Multimodal systems encompass more than just LLMs. For instance, text-to-image models such as Midjourney, Stable Diffusion, and Dall-E are multimodal but lack a language model component. The term "multimodal" can refer to various scenarios, including different modalities for input and output (text-to-image, image-to-text), processing of multimodal inputs (text and images), or generation of multimodal outputs. Therefore, fine-grained taxonomies of FMs could be useful, as well as variation of corresponding technological stack and mitigation measures.

Open artificial intelligence systems and products that are being served to users can integrate more components than what is strictly needed to build, document and distribute open models. For instance, to build an open chatbot system, developers might add an external content moderation system (or other types of safety guardrails) on top of the open model, build a UI and collect usage telemetry and logs. Similarly, when developing a RAG (Retrieval augmented generation) product for a search application, a developer may add an orchestration layer (langChain, llamaIndex), or a LLM cache (Redis, SQLite), etc.

AI systems are composed of various components.[9] The AI system stack is schematically summarized around a layer of components relevant to model itself, along with the infrastructure on which it relies (below), and the Product / UX that may be built on top of it (above).

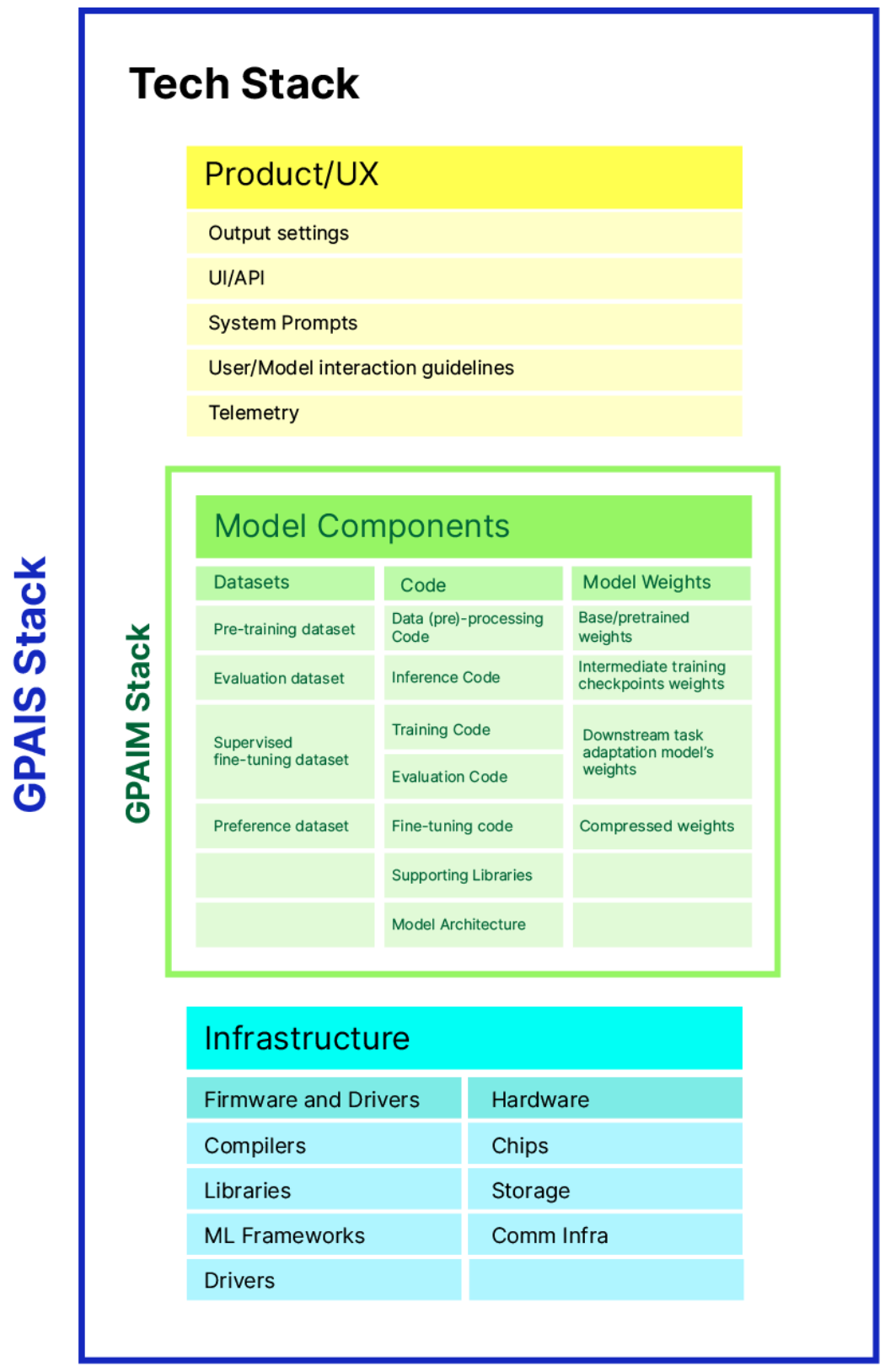

**Fig 3: General Purpose AI System (GPAIS) Tech Stack**

This representation underlines the differences between AI systems and traditional software. With traditional software, there was a very clear separation between the code written, the compiler used, and license possessed. One could use an open source compiler to write proprietary code, or a closed source integrated development environment (IDE) and compiler and write open source code. Also, when inspecting traditional software systems, access to source code could facilitate explaining and reproducing behaviors.

However, for AI models and systems, many components collectively influence the functioning of the system, including the algorithms, code, hardware, and datasets used for training and testing. The very notion of modifying the source code (which is important in the definition of open source) becomes fuzzy. For example, when considering FMs[10] , should the dataset be considered as the source code for the model/weights that have been trained? Should the model itself be considered the preferred artifact for making modifications to an AI system? These factors are not necessarily observable by inspecting the final AI system. In addition, the presumed benefits of one artifact over another are not always true — for example, if a model has been trained with external data, there does not seem to be a trivial way to remove the piece of information (if it proves fraudulent, or if consent to use has been lacking or revoked, for example) from the model itself without retraining the whole system. One might ordinarily assume that the model/weights might constitute better artifacts for preserving privacy, but that is not necessarily true for AI models.[11]

[8] Advanced FM models (alternatively, general purpose AI models with systemic risk) define the most capable models whose performances match or outperform existing models either in terms of capabilities or variety of tasks. They are already subject to particular scrutiny from regulators. The EU AI Act and the U.S Executive Order on AI of October 2023 define those models with a compute threshold (respectively any training run involving over $10^{25}$ and $10^{26}$ floating-point operations is covered).

[9] Various granular representations have been suggested by organizations and academia, including the Ada Lovelace Institute, the AI Centre for the governance of AI, Stanford AI Index, Open Future, Creative Commons.

Lastly, the infrastructure layer plays a key part, as many scholars already underlined: the high costs associated with developing AI models could hinder widespread adoption, even using open approaches, due to the expensive hardware and infrastructure requirements like GPUs. While openness in AI provides organizational control for customization, adapting large models to specific domains poses challenges without substantial investments in GPUs, so this does not necessarily solve the challenges of concentrated access and influence in well-resourced labs.

[10] Examples of FMs include large-scale autoregressive language models like decoder-only or encoder-decoder models, as opposed to encoder-only models that are used to embed text for various downstream NLP tasks such as classification but are not text generators; diffusion models, text-to-image models etc.

[11] Despite these differences between building AI models and traditional software, some initiatives are trying to study to what extent foundation models could be built “as open source software” — especially given the technical ability to efficiently update the weights of the models. This would allow contributors to be able to submit patches to fix potential identified vulnerabilities in the model, and then maintainers would need to be able to merge the patches to quickly integrate updates into the model.

### 3.3. Common attributes across the AI system stack

It is crucial to differentiate between the components of the tech stack (as described in section 3.1. and 3.2.), such as model weights, and the attributes associated with these components, such as the license for model weights. Indeed, some frameworks may overlook this distinction. For example, we note that safety measures are better viewed as an attribute of the system rather than a feature of the model.

Consequently, we represented an AI system amidst three cross-cutting sets of attributes, which may themselves be attached to other components across the whole AI system: documentation, safeguards and licensing (**Figure 4**).

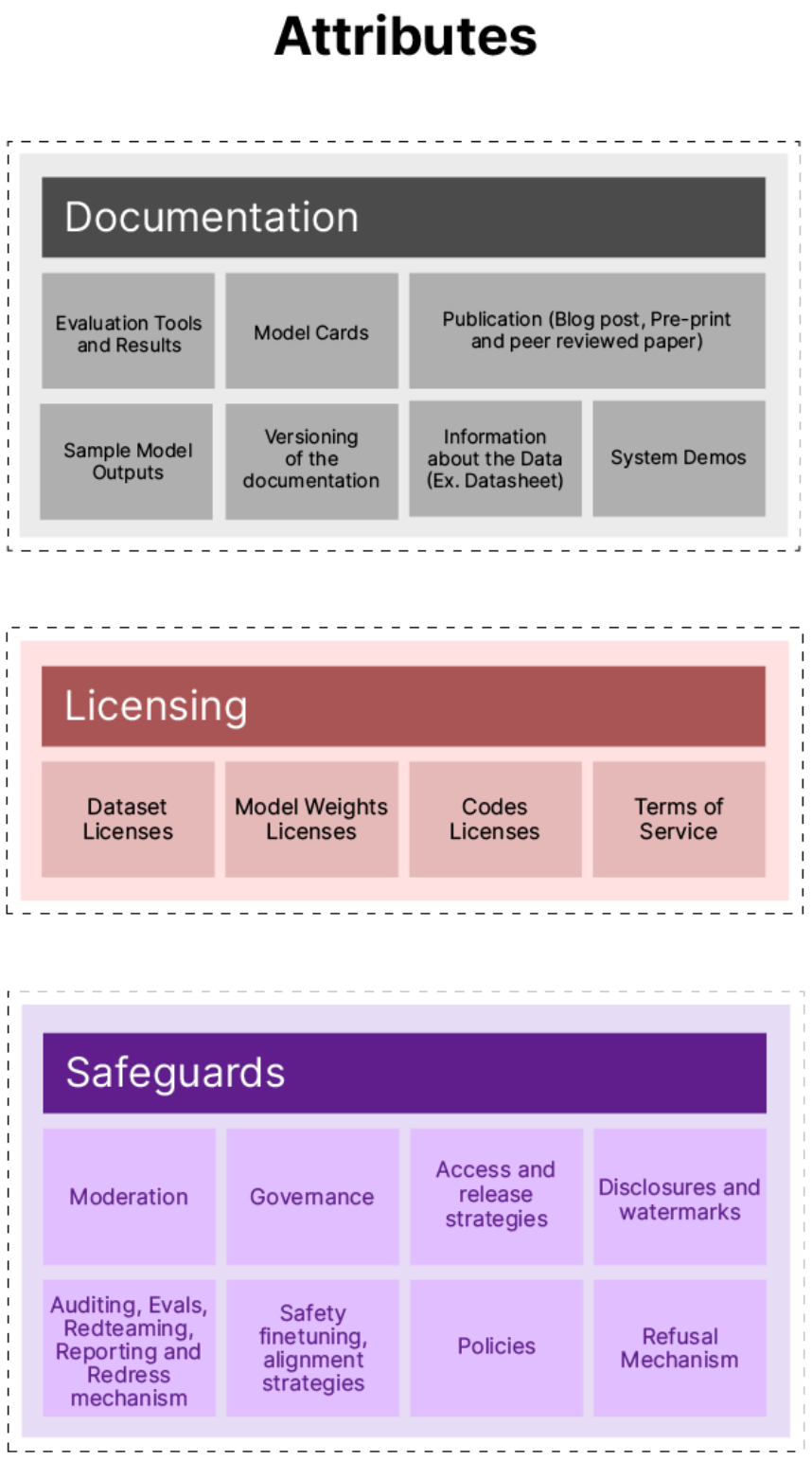

**Fig 4: Cross-set of attributes across the whole GPAIS Stack**

Appendix I presents some of these attributes. To illustrate, safety guardrails could be integrated at the level of each component, to acknowledge their diversity to tackle different types or harms and risks emerging throughout the stack, and the existing mitigations best suited to handle them. As one article notes, “AI safety questions cannot be asked and answered at the levels of models alone. Safety depends to a large extent on the context and the environment in which the AI model or AI system is deployed.”[12] Entrusting safety assessments solely to models would be insufficient, especially regarding concerns of potential misuse, as models lack crucial contextual information necessary for making accurate safety assessments.[13]

### 3.4. An Openness in AI Framework

As a whole, the framework presents a tech stack described across its different dimensions: AI artifacts (e.g. model’s weights), to which correspond different attributes such as licensing (e.g. choice of model’s weights license) or documentation (e.g. model cards).

Each of the dimensions are composed of categories and subcategories depending on the relevant level of granularity one may choose to consider. Each of these categories (or subcategories) can be released more or less openly.

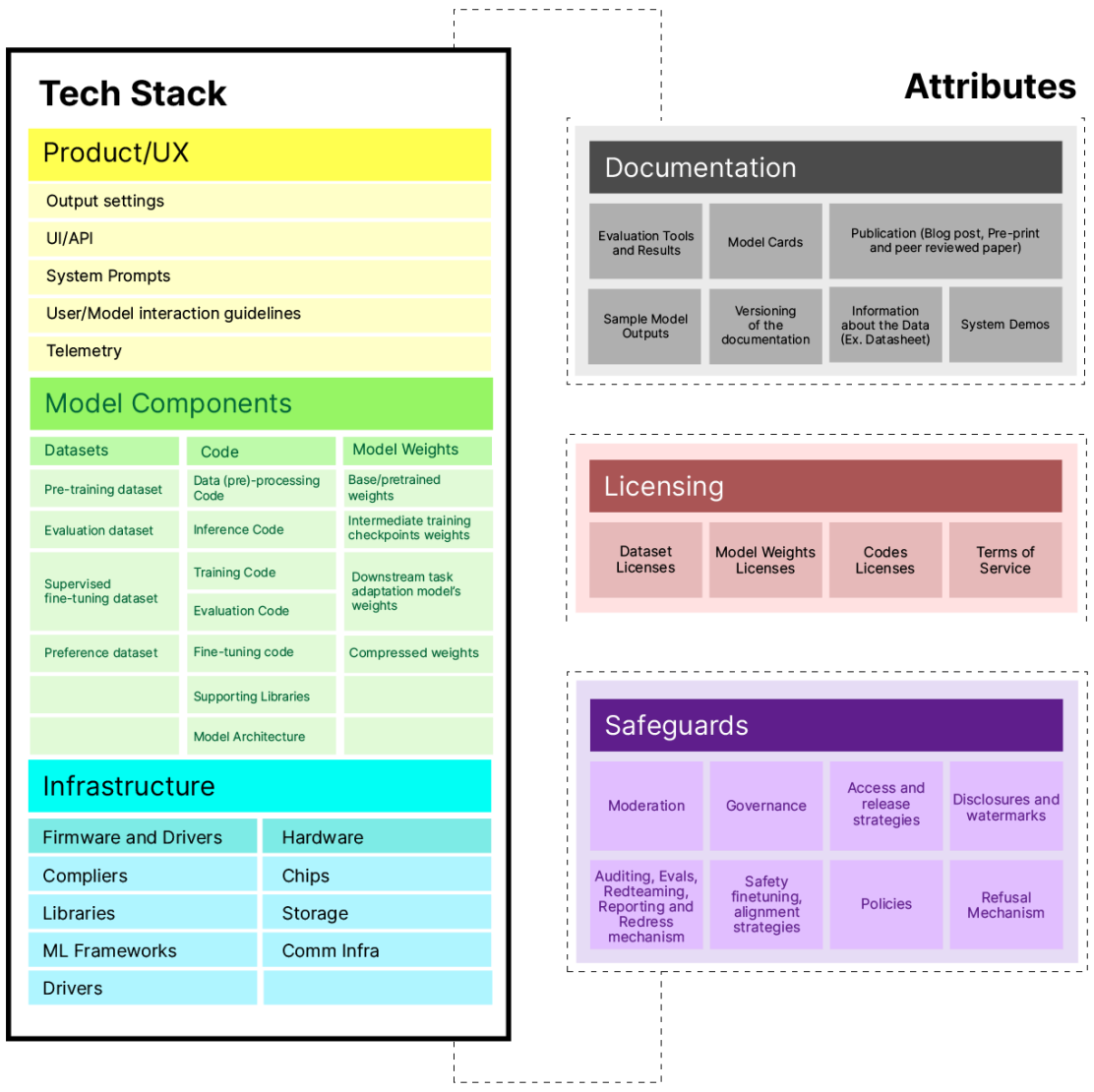

**Fig 5: Dimensions of AI Openness Framework**

Granting access to each of these categories or sub-categories of technological stack enhances various purposes. Appendix 1 surveys some of these components to discuss different benefits of openness at each layer of the stack.

Further work would need to be undertaken, notably to answer questions like: What do people and organizations need to make various components of the stack open? For builders and developers, what would be the minimum components to ensure reproducibility vs replicable evaluation? More generally, what subset of components matter for what kind of use case? How do we future proof these analyses for the next generation of AI systems (e.g., those that are more objective-based than LLMs)?

### 3.5. Why pursue AI openness?

Some people disagree on the primary goal(s) of AI openness. Though these motivations are not mutually exclusive, often one does not pursue AI openness for openness’ sake. In **Fig. 6** below is a brief overview of the main benefits that can be achieved through AI openness.

[12] Arvind Narayanan and Sayash Kapoor, “AI safety is not a model property,” AI Snake Oil, Mar. 2024.
[13] Certain types of content inherently pose risks, such as AI-generated non consensual intimate imagery. Rather than relying on red teaming to determine whether a model can be misused, its utility lies in understanding the evolving landscape of adversary capabilities facilitated by cutting-edge AI models.

1. Enhance reproducible research, promote innovation (open science)
2. Share knowledge, educate people (open knowledge)
3. Create an ecosystem of developers, makers, builders, foster collaboration (open ecosystem)
4. Create building blocks, infrastructure that people can extend to create things, advance the field (accelerating innovation)
5. Give the tools so that people can train on various sources/datasets for diversity of representation and visions (open culture, models behavior)
6. Ensure accountability of systems, via audits, explainability of systems
7. Support research on bias, environmental and other key challenges (equality, non discrimination)
8. Foster the unexpected or the unimagined
9. Identify bug, security defaults (widespread scrutiny & security)
10. Reduce cost for (public) procurement (cost optimization)
11. Avoid vendor lock-in (increase digital autonomy) enabling more private (e.g. local) architectures as opposed to cloud-based
12. Independent researchers, regulators need tools to assess large scale AI systems for risks (equip supervising authorities adequately)
13. Make training and inference more resource-efficient (environmental impact)
14. Ensuring competition dynamism in the field, balance the dominant positions by new entrants and smaller players (competition)
15. Ensuring recourse. When decisions impact humans negatively or incorrectly, in closed systems there is no recourse.

**Fig 6: Examples of Potential Benefits from Openness**

This list is not exhaustive. It shows the variety of motivations and purposes of AI openness. It also shows that even if AI openness is challenged on one aspect (e.g. security), it could still play an important role on another aspect (e.g., competition). This also suggests that a multidimensional approach to AI openness could be constructive to underline the various possibilities, including how different levels of access enable specific benefits of openness.

## 4. Beyond "Open Bashing" and "Open Washing"

Over the past two years, a lot of foundational work has been done to map the space of openness in AI. Below are a selection of key papers that suggest different approaches for mapping dimensions of AI openness. These papers are sorted based on whether they characterize openness as more of a gradient ("more vs. less open"), a score ("certain amount of open"), or a binary ("is open or closed"). A list of controversial topics that are about or adjacent to dimensions of openness is available in Appendix II.

### 4.1. Gradient / Spectrum approaches

*Characterizing openness on a gradient with different levels of openness.*

The Gradient of Generative AI Release: Methods and Considerations (I. Solaiman, Feb 2023)

This paper introduces gradients of generative AI releases. The abstract sums the paper best: "As increasingly powerful generative AI systems are developed, the release method greatly varies. We propose a framework to assess six levels of access to generative AI systems: fully closed; gradual or staged access; hosted access; cloud-based or API access; downloadable access; and fully open. Each level, from fully closed to fully open, can be viewed as an option along a gradient. We outline key considerations across this gradient: release methods come with tradeoffs, especially around the tension between concentrating power and mitigating risks."

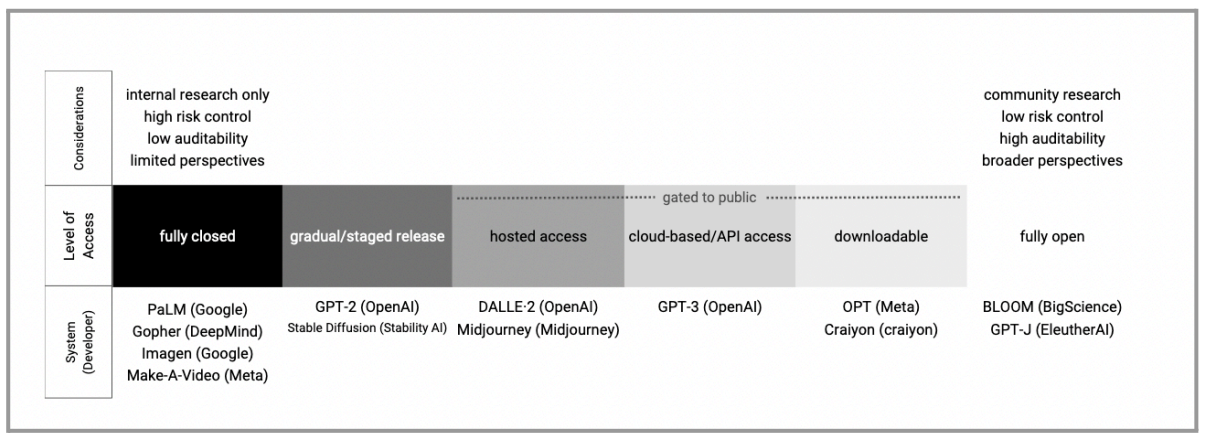

**Fig 7: Considerations and systems along the gradient of system access**

Workshop on Responsible and Open Foundation Models (open foundation models, Sept. 2023)

In September 2023, Princeton Center for Information Technology Policy and Stanford Center for Research on Foundation Models hosted a workshop to discuss key benefits and risks of specific properties of open FMs . They proposed a risk assessment framework for analyzing their *marginal risk* (i.e. using closed-source and pre-existing technologies as a baseline). They identify that more research is needed to assess the marginal risks of open FMs.

Gradient Approach to the Openness of AI System Components (Digital Public Goods Alliance - UNICEF, Oct. 2023)

This gradient approach framework recognizes that having access to AI training and testing data does not equate to having access to source code. This implies that the mere act of open-sourcing the training data used does not enable the recreation of the model. It thus recognizes varying degrees of openness in AI systems, especially regarding training data, model, and code, without further detailing sub-components. The aim is to balance the benefits of AI openness, like transparency and inclusivity, against risks like misuse or harm. This framework includes the concept of "aspirational openness," defining various levels of openness for AI components to qualify as digital public goods, taking into account factors like ethical use and data rights issues.

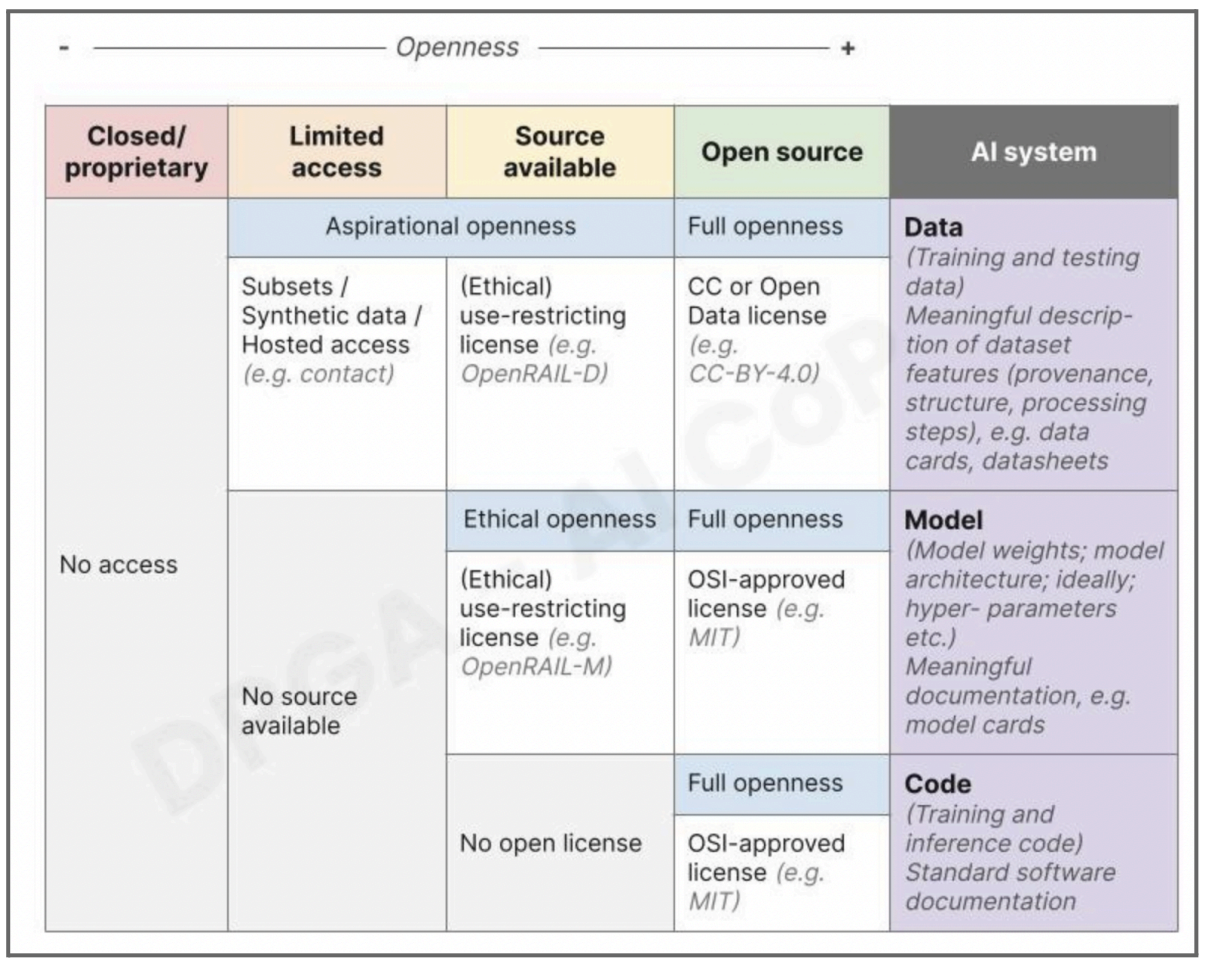

**Fig 8: Tracking openness across components of an AI system**

[Triangle Badges] (link) (Tweet, Biderman, Dec. 2023)

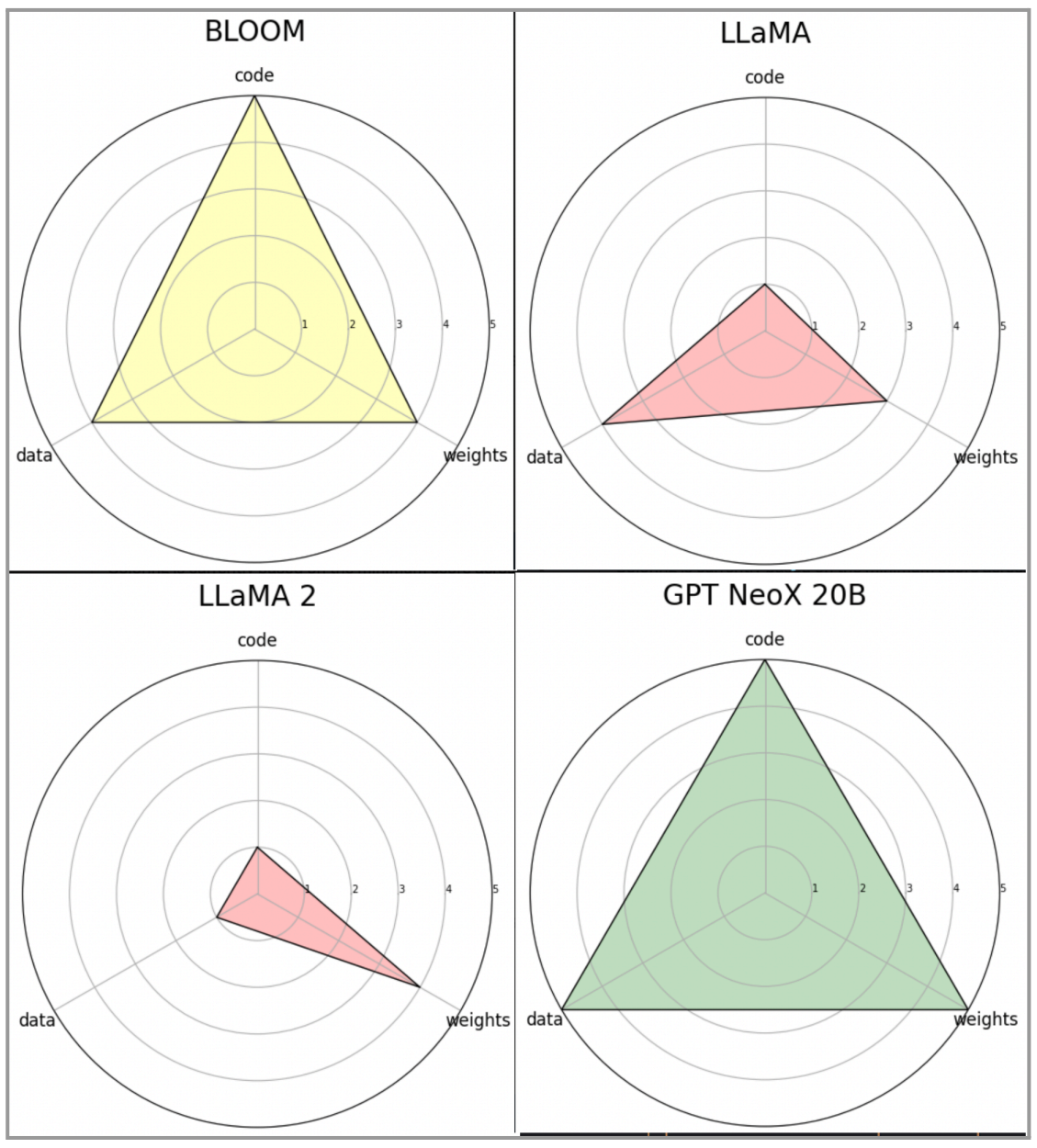

**Fig 9: Tweet: Radar charts visualizing different dimensions of openness (code, data, weights)**

Considerations for Governing Open Foundation Models (R. Bommasani, S. Kapoor, K. Klyman, S. Longpre, A. Ramaswami, D. Zhang, M. Schaake, D. Ho, A. Narayanan, P. Liang, Dec. 2023)

This brief discusses how release strategies of open foundation models (i.e. "to what extent and through what mechanisms are foundation models made available to entities beyond the foundation model developer?") relate to governance risks and regulation consequences. The paper builds upon Irene Solaiman's above graphic breaking down the "downloadable" and "API access" categories into more granular components. A key takeaway of the brief is to invite practitioners to think in terms of *marginal risk of open foundation models* "relative to (a) closed models or (b) pre-existing technologies." The paper also discusses some of the benefits of open foundation models like competition, accelerating innovation and distributing power. The brief acknowledges that the release of foundation models is a gradient across various dimensions, and encourages policymakers to explore downstream choke points that would be more effective for mitigating AI risks.

On the Societal Impact of Open Foundation Models (S. Kapoor, R. Bommasani et al., Feb. 2024)

The paper presents an analysis of the risks and benefits of open foundation models (with widely available weights). The paper analyzes benefits spanning innovation, competition, the distribution of decision-making power, and transparency. It also presents a risk assessment framework for evaluating the marginal risk of open foundation models (following the recommendation in the policy brief), with six steps: (i) clearly specifying the threat model; (ii) Analyzing existing risk absent open FMs; (iii) Analyzing existing defenses (absent open FMs); (iv) presenting evidence or an analysis of the marginal risk of releasing models weights; (v) Analyzing the difficulty of defending against this marginal risk; (vi) Articulating the uncertainty and assumptions in the process of conducting this analysis. In addition, the paper looks at a variety of work on the risks of open foundation models and in many cases, finds current evidence lacking.

### 4.2. Criteria Scoring Approach

*Providing a score for openness based on different attributes.*

[“Opening up ChatGPT”](#) (A. Liesenfeld, A. Lopez, M. Dingemanse, July 2023)

This project by the Centre for Language Studies at Radboud University ranks the openness, transparency, and accountability of instruction-tuned language models. It reviews risks associated with proprietary software and surveys open-source projects with similar architecture and functionality. The project evaluates AI models based on the 13 criteria: Availability (is it closed VS partially open VS fully open for inspection?) for : Source code, LLM training data, LLM model weights, RLHF training data, RLHF model weights, License (by reference to a OSI approved license). Documentation (is it available and fully documented VS partially documented?) for: Code, System architecture and model training setup, Preprints (archived preprints that cover all parts of the software including base models, fine-tuning, and RLHF components), Papers (peer-reviewed papers that cover all parts of the software including base models, fine-tuning, and RLHF components), Model Cards, Data Sheets. Access methods in relation to: Index Package Software (A packaged release of fully open-source software), API (commercial-restricted access vs. open API). Many tools built on existing large language models inherit the undocumented datasets these base models are trained on.

| Project | Availability | | | | | | Documentation | | | | | | Access | |
|---|---|---|---|---|---|---|---|---|---|---|---|---|---|---|
| (maker, bases, URL) | Open code | LLM data | LLM weights | RL data | RL weights | License | Code | Architecture | Preprint | Paper | Modelcard | Datasheet | Package | API |
| BLOOMZ | ✔ | ✔ | ✔ | ✔ | – | – | ✔ | ✔ | ✔ | ✔ | ✔ | ✔ | ✘ | ✔ |
| bigscience-workshop | LLM base: BLOOMZ, mT0 | | | RL base: xP3 | | | | | | | | | | § |
| Open Assistant | ✔ | ✔ | ✔ | ✔ | ✘ | ✔ | ✔ | ✔ | – | ✘ | ✘ | ✘ | ✔ | ✔ |
| LAION-AI | LLM base: Pythia 12B | | | RL base: OpenAssistant Conversations | | | | | | | | | | § |
| AmberChat | – | ✔ | ✔ | ✔ | – | ✔ | – | – | ✔ | ✘ | ✔ | – | ✘ | ✔ |
| LLM360 | LLM base: LLaMA-7B | | | RL base: Evol-Instruct (synthetic) | | | | | | | | | | § |
| Pythia-Chat-Base-7… | ✔ | ✔ | ✔ | ✔ | ✘ | ✔ | ✔ | ✔ | – | ✘ | – | – | ✔ | ✘ |
| togethercomputer | LLM base: EleutherAI pythia | | | RL base: OIG | | | | | | | | | | § |
| RedPajama-INCITE… | – | ✔ | ✔ | ✔ | ✔ | – | – | – | ✘ | ✘ | ✔ | ✔ | ✘ | – |
| TogetherComputer | LLM base: RedPajama-INCITE-7B-Base | | | RL base: various (GPT-JT recipe) | | | | | | | | | | § |

**Fig 10: Evaluation of openness across instruction tuned LLMs (cropped)**

[The Foundation Model Transparency Index](#) (Stanford CRFM and HAI, Oct. 2023)

This Index specifies 100 fine-grained indicators that codify transparency for FMs, from the upstream resources (e.g. data, labor, compute), the model itself (size, capabilities, risks), and downstream uses (distribution channels, usage policies, affected geographies). AI release strategies exhibit a non-binary nature. This study classifies models with widely downloadable weights as falling within the open category. Importantly, models labeled as open showcase superior performance. Many of these disparities are ascribed to the closed developers' lack of transparency regarding critical upstream matters, such as data, labor, and computational resources used in model construction. EleutherAI, in response to this paper, later highlighted a different [perspective](#) on transparency, notably raising that score-based frameworks can lead to transparency being viewed as an overarching goal to achieve, instead of a tool or "a mechanism for achieving other ethical values, such as accountability."

| Major Dimensions of Transparency | Meta Llama 2 | BigScience BLOOMZ | OpenAI GPT-4 | stability.ai Stable Diffusion 2 | Google PaLM 2 | ANTHROPIC Claude 2 | cohere Command | AI21labs Jurassic-2 | Inflection Inflection-1 | amazon Titan Text | Average |
|---|---|---|---|---|---|---|---|---|---|---|---|
| Data | 40% | 60% | 20% | 40% | 20% | 0% | 20% | 0% | 0% | 0% | 20% |
| Labor | 29% | 86% | 14% | 14% | 0% | 29% | 0% | 0% | 0% | 0% | 17% |
| Compute | 57% | 14% | 14% | 57% | 14% | 0% | 14% | 0% | 0% | 0% | 17% |
| Methods | 75% | 100% | 50% | 100% | 75% | 75% | 0% | 0% | 0% | 0% | 48% |
| Model Basics | 100% | 100% | 50% | 83% | 67% | 67% | 50% | 33% | 50% | 33% | 63% |
| Model Access | 100% | 100% | 67% | 100% | 33% | 33% | 67% | 33% | 0% | 33% | 57% |
| Capabilities | 60% | 80% | 100% | 40% | 80% | 80% | 60% | 60% | 40% | 20% | 62% |
| Risks | 57% | 0% | 57% | 14% | 29% | 29% | 29% | 29% | 0% | 0% | 24% |
| Mitigations | 60% | 0% | 60% | 0% | 40% | 40% | 20% | 0% | 20% | 20% | 26% |
| Distribution | 71% | 71% | 57% | 71% | 71% | 57% | 57% | 43% | 43% | 43% | 59% |
| Usage Policy | 40% | 20% | 80% | 40% | 60% | 60% | 40% | 20% | 60% | 20% | 44% |
| Feedback | 33% | 33% | 33% | 33% | 33% | 33% | 33% | 33% | 33% | 0% | 30% |
| Impact | 14% | 14% | 14% | 14% | 14% | 0% | 14% | 14% | 14% | 0% | 11% |
| Average | 57% | 52% | 47% | 47% | 41% | 39% | 31% | 20% | 20% | 13% | |

**Fig 11: Foundation Model Transparency Index Scores by Major Dimensions of Transparency, 2023**

### 4.3. Binary approaches

*Characterizing openness as a binary — it's either open or closed.*

[Defining Open Source AI](#) (OSI, ongoing), [Deep Dive AI](#) (OSI, Feb. 2023) and [Llama-2-license is not open source](#) (OSI, July, 2023)

In June of 2023, OSI launched a process to draft a definition of open source AI. The process included an open call for papers and a webinar series covering the landscape of OS AI, commons based data governance, Data Privacy and operationalizing openness in AI and others. The process continues in 2024 with a public conversation and a regular release of drafts until an expected release of a usable version 1.0 of the Open Source AI Definition by October 2024.

Pythia: A Suite for Analyzing Large Language Models Across Training and Scaling (EleutherAI, May 2023)

EleutherAI launched the pioneering effort of releasing a suite of fully open models "providing public access to 154 checkpoints for each one of the 16 models, alongside tools to download and reconstruct their exact training dataloaders for further study". In a tweet on December 2023, Executive Director of EleutherAI promotes open source licenses arguing "it fosters interoperability among codebases, clarifies the legal status of the use of code and protects both the software owner and the software user from frivolous claims" while emphasizing that custom licenses hurt the ecosystem.

LLM360: Towards Fully Transparent Open-Source LLMs (LLM360, Dec 2023)

LLM360 initiative aims at "open-sourcing LLMs from all angles": "all training and model details (e.g., hyperparameters, schedules, architecture, and designs), all intermediate model checkpoints saved during training, and full disclosure of the exact pre-training data used". For LLM360, releasing partial artifacts hinders progress in the field by degrading transparency into the training of LLMs and forcing teams to rediscover many details in the training process.

OLMo: Accelerating the Science of Language Models (AI2, February 2024)

The Allen Institute for AI built on the likes of Pythia and BLOOM by releasing an initial set of 1B and 7B models with 500+ checkpoints from intermediate training, full datasets, training code, evaluation code, training logs, and everything in between. The goal of this initiative beyond further transparency and openness is to foster a growing "science of LLMs" driving the public narrative of AI.

The Model Openness Framework: Promoting Completeness and Openness for Reproducibility, Transparency and Usability in AI (M. White, I. Haddad, C. Osborne , Xiao-Yang (Yanglet) Liu, A. Abdelmonsef, S. Varghese, March 2024)

The Generative AI Commons, an open participation group at the LF AI & Data Foundation has developed the Model Openness Framework (MOF). The MOF is a 3-tiered classification system that ranks the openness and completeness of machine learning (ML) models. It is built on principles of open science, open source and open data, and requires content-appropriate open licenses to be used for every artifact that is included in a model's distribution. There are 15 components in the framework that come from the model development lifecycle and those artifacts are associated with different classes of the MOF. In addition to the MOF components, the framework requires the inclusion of a master inventory file, MOF.JSON, which includes information on the components included and their corresponding open licenses. The MOF classes range from Class III (Open Model) which includes the minimum required components for model usage, to Class II (Open Tooling), and to Class I (Open Science) which is the gold standard in the MOF, where all components are available using open licenses. The goal of the MOF is to encourage AI researchers, developers and organizations that produce ML models to include as many components as possible and to employ open licenses. This has the benefit of maximizing transparency and reproducibility for ML models and allows them to be studied, modified, redistributed and used for any purpose including commercial applications without restrictions.

**Conclusion**

This work aimed to accelerate a shared understanding of openness in FMs and AI systems, differentiating challenges at the model level, the system level, and attributes that could apply across the AI stack, and how to leverage that understanding to better harness the benefits and mitigate the risks of AI. In future work, we hope to go deeper into how this framework can be strengthened with more information about safety mitigations.

**Acknowledgements**

We acknowledge Aviya Skowron, Amba Kak, Irene Solaiman, Stella Biderman and Martin Tisné for their input to this paper. We also thank all the participants in the Columbia Convening on Openness and AI for their contributions to this work.

## APPENDIX I - Opening the foundation model stack

The illustrations below are not exhaustive, and aim at pointing out which purposes of openness in AI can be pursued. Increasing access to some of these categories or sub-categories of technological stack can enhance reproducible research, auditability, and transparency. Although one can benefit from accessing each of the resources individually, new benefits can be revealed when accessing several resources together (e.g., accessing datasets, a model's architecture as well as data quality checks can help establish compute-optimal laws like "chinchilla"). Clarity is paramount when making distinctions to clearly identify the component under discussion and explicitly define its intended use.

### AI MODEL STACK

#### DATASETS

- **Pre-training datasets:** Datasets, often very large datasets that are snapshots of the web, on which models are trained. Examples include Pile, C4, ROOTS, RefinedWeb, Dolma. More openness can enable one to study the biases, fairness, toxicity of the data on which the model is trained; open datasets such as CommonCrawl which plays a central role in training of LLMs today, enabling researchers to document hate speech and low quality content that is fed to the models. Privacy-preserving approaches can be particularly important here, in order to properly balance considerations around security and confidentiality (e.g, in health).
- **Supervised fine-tuning datasets:** Smaller, context-specific datasets than are used to fine-tune the model on specific tasks and domains, such as instruction-tuning datasets and dialogue datasets. More openness can allow one to understand how the model was specialized for a specialized downstream task (e.g. dialogue, coding assistant).
- **Preference datasets:** Datasets that compare and classify model outputs on conversation text data. This can include RLHF datasets and human preference datasets, which play a central role in aligning models on user preferences, content moderation, style, etc. More openness can enable better language and image modeling to understand how the model was aligned, and more transparency on content moderation and safety approaches.
- **Evaluation datasets:** Test datasets that are used to evaluate the models' performance offline and derive various metrics on top of them. More openness can increase the transparency of data on which the model was evaluated (which can be tricky to do reliably since base pre-trained models are general purpose) and enable auditing by third-parties including end users. More openness can also enable researchers to identify and fix gaps in evaluation procedures, benchmarks, and scenarios.
- **Evaluation prompts:** Pieces of text that encapsulate samples of text of evaluation datasets with additional guidelines or context (e.g., "You are an expert in legal, please answer the following question + [Insert the evaluation sample]"). More openness can enable better reproduction of evaluation results, and fairer comparison across models as spacing, punctuation, and casing can currently have an important impact on LLM performance.

#### CODE

Granting access to each of this category of code empowers developers and researchers differently. For instance:

- **Data (pre)-processing code:** Code that is applied to the dataset before it is passed to the foundation model, such as tokenizers, data deduplication code, data quality checks, and various filters. More openness can enable better assessments of the quality of data; provide information about data deduplication, data quality, and fairness; and improve our understanding of how data processing influences different dimensions such as performance and privacy.
- **Pre-training code:** Code that is used to train the foundation model, which defines the training cycles. It includes the model's architecture, training loss, hyperparameters and configurations. More openness can enable more people to participate in retraining the model on other datasets and to understand how different training strategies influence different training objectives (e.g., assessing how different training strategies impact memorization, and what one should do at training time to build a more robust model)
- **Training libraries:** Pre-built scripts that are imported by the developer and used as packages during the training of the foundation model. Examples include Transformers, Keras, GPT-NeoX, and llm-foundry. More openness can enable more people to participate in retraining the model on other datasets, and is vital for open science and open knowledge approaches (e.g., reproducibility).
- **Fine-tuning code and libraries:** Code that is used to fine-tune a model on a particular domain or application. This can include methods like SFT, RLHF, and DPO, and fine-tuning libraries like alignment-handbook, LoRA and PEFT. More openness can lower costs for a broader community of researchers to build and reuse more efficient fine-tuned models, produce thriving ecosystems of task-specific models built on top of pretrained models, and enable safety researchers to scientifically advance risk mitigation techniques (e.g., alignment methods, preference sampling training methods).
- **Inference code:** Code that enables the models to serve results to the users. This includes prompting design, inference optimization, early predictions, decoding methods, and search. More openness can allow developers to better integrate the model into downstream applications, and open source frameworks (e.g., vLLM, FlashAttention, BitsandBytes, PowerInfer, and techniques like speculative inference) have helped make inference cheaper and models more efficient.
- **Distributed computing libraries:** Libraries that contain the code that is used for training the models on distributed clusters (multi GPUs, multi nodes). This includes SOTA training techniques such as distributed training, mixed precision, and gradient accumulation, and libraries such as Accelerate, DeepSpeed, and Pytorch. More openness via increased access to base distributed training libraries can enable more efficient training on different clusters, and empirically assessing model and data parallel training techniques can enable one to better understand memory footprint and computational efficiency.
- **Inference and cloud infrastructure frameworks:** The software infrastructure that is used during inference. This includes approaches such as scalable serving, multi-cloud optimization, and batching, and frameworks and tools such as SkyPilot, Ray Serve, Kubernetes, vLLM, and Triton. More openness can enable better reproducibility and enables various degrees of interoperability and portability of the code on different types of hardware and clouds.

**MODEL WEIGHTS**

- **Pretrained weights:** Numerical values of all the parameters of models that have been tuned during the self-supervised pre-training phase. Examples include Mistral-7B, llama-2-70B, and Pythia-12B. More openness can enable developers to build an ecosystem of developments on top of one model (e.g "the stable diffusion moment"), repurpose models for different tasks, train models in ways that are more attuned to linguistic and cultural attributes around the world, and enable more auditability, scrutiny, privacy, and transparency. It can also foster entire new research avenues like mechanistic interpretability that require access to the internals of the model, and it can help with comparing architectures' efficiencies to improve the scientific understanding on the generalization of models.
- **Intermediate training checkpoints' weights:** Intermediate values that the parameters of the model can take during the training phase; at regular timesteps, those values can be saved as intermediate models. More openness can enable the community of researchers to pursue scientific research on scaling (and inverse scaling) laws that are critical to understand the training dynamic of the technology.
- **Downstream task adaptation of model weights:** Weights of fine tuned models that have been specialized for a specific task or domain. This includes instruction fine-tuning, alignment, supervised fine-tuning, etc. More openness can enable developers to seamlessly integrate highly-capable models into user-facing systems, enable third-party auditors to provide more scrutiny to these models, and enable researchers to study the safety costs associated with such custom fine-tuning. Releasing aligned models enables researchers to better evaluate how robust and safe alignment methods are currently, and to pursue new research avenues like research on universal jailbreak backdoors on aligned models.
- **Compressed and adapter weights:** Weights that result from the compression of a model — the process of making a model smaller while not compromising on performance. This includes processes like pruning and quantization, and adapter weights like LoRA and QLoRA. More openness can enable the community to better study inverse scaling laws with more powerful small models, enable quick model adaptation (as sharing LoRA weights is very convenient) and open new research avenues in models' efficiency.
- **Reward model weights:** Weights resulting from models that are learned from preference datasets to enable automatic classification or to rank models' text output according to users' preferences. These models are further used to fine-tune a foundation model based on users' preferences. More openness can enable sharing of models that have been trained on human or AI preferences, which can reduce the barrier to using safer models in practice. It can also help other AI models follow instructions better, and those reward models can also be further fine-tuned by developers for custom needs or for further rejection sampling training.

**ATTRIBUTES**

This paper suggests that thorough considerations of openness in AI systems will consider the wide set of attributes that may be opened across the stack, and across the three key areas of safeguards, documentation, and licensing.

Each component of the GPAIM stack can have various attributes, depending on the types of release and licenses associated with it. Here are some illustrations (not exhaustive):

## DOCUMENTATION

Currently, FMs are often not documented with high levels of granularity, and the specifics of documentation vary from project to project. The following items are a list of attributes that can be included as part of documentation, and explaining the reasons why some information is not shared may also be a helpful contribution to openness.

### DATA DOCUMENTATION

- **Dataset characteristics:** Metadata informing content of a dataset, such as task distribution, language distribution, topics, format, etc. More openness can enable developers to get detailed information relevant to understanding data representation and data composition in order to keep control on the nature of the data on which a model has been trained.
- **Data provenance:** Metadata relative to the provenance and history of the dataset like versioning, attribution, text source, citation and download counts. More openness can enable developers to keep track of the history, versions and sources of the datasets especially when datasets are iteratively aggregated and merged overtime.
- **Data annotation, third-party annotators & labelers:** General guidelines that have been given to labelers to capture designers' intent in shaping an optimization (e.g., when crafting the data instructions / filtering practices for iterative/online RLHF). More openness can enable developers to understand possible preference biases, how labels have been assigned, as well as potential outsourcing labor problems and discrimination.
- **Data quality checks and qualitative analyses:** A series of tests informing about the quality of the dataset (e.g. semantic and linguistic checks). More openness can inform developers about the potential downstream vulnerabilities of the models.

### MODEL CARDS

- **Intended use:** Attributes like model objectives and out of scope use cases that inform downstream developers about the expected use of the model when originally designed by the model's provider. More openness can enable downstream developers to understand the design choices and the benefits and risks of using the models for various use cases and contexts.
- **Model technical details:** Attributes about the design of a model, such as model's architecture, hyperparameters, and tradeoffs in design choices. More openness can enable developers to get contextual information about when to use the models and to get a better sense of the model providers' intentions.
- **Compute resources:** Information on compute resources that allow a broad spectrum of stakeholders to understand what amount of compute was used as it relates to hardware/software efficiency, carbon footprint of AI and portability. Examples include energy efficiency, GPU/hardware specification, amount of compute and time required to run inference on fixed hardware for a specific task. More openness can make it easier to design more efficient models and track the carbon footprint of AI, both from training and inference.

- **Evaluation:** Documentation of evaluation results and protocols, which can enable people to understand the capabilities and limitations of a model. This can include qualitative and quantitative evaluation protocols, simulation environments, metrics and results, prompting techniques, evaluation tasks, and benchmarks. More openness can enable the community to challenge existing state-of-the-art models while conducting fair comparison across models across different metrics (e.g., fairness, robustness, efficiency) and enable developers to understand strengths and vulnerabilities of the produced models. Notably, newly surfaced vulnerabilities in AI products revealed during red team testings. More openness can help downstream developers better understand the risks associated with the deployed model, and it can help increase foresight into how the model could cause harm when used by the general public or by malicious actors.

**PUBLICATIONS**

- **Pre-print and peer reviewed paper:** A paper that is accessible online before it has gone through the peer-review process of a scientific conference or a journal. More openness (i.e., more pre-prints and peer-reviewed papers) is a base practice to advance the scientific discipline of AI, and in the recent history of AI, scientific papers have been published at open access conferences, journals, and websites like ArXiv and OpenReview.

- **Impact Assessment & red teaming reports:** Impact assessments are structured processes to imagine, document and quantify the possible impacts of a proposed AI system on various stakeholders. More openness can increase public awareness and transparency about document downstream use and risks of a model as well as the societal implications of the release. The question arises on the possibility for open foundation models, that are general purpose by design and for which the distribution of users might not be fully known. (See, e.g., the AI-Risk Management Standards Profile for GPAIS and FM.)
- **System demos:** Toy live user interfaces where users can play with a given AI prototypes; they can usually handle a large workload (e.g., Hugging Face spaces or the openplayground). More openness can enable the public to test models in real time, helping increase public awareness and familiarity with the technology and develop a better mental model of it — which can help increase earned trust in AI systems. It also helps developers run qualitative evaluations and one-to-one comparisons of models' outputs, and it can facilitate red-teaming efforts and enable scrutiny from stakeholders with less technical resources.

**LICENSE**

- **License(s) applicable to data:** Legal documents that define the use and access of data through a contractual agreement between the data provider (licensor) and the data user (licensee). Examples include the Linux Foundation's Community Data License Agreement (CDLA-Permissive) and the Open Data Commons licenses like Public Domain Dedication and License (PDDL) and the Open Data Commons Attribution License License (ODC-By). More openness via increased access to data licenses can enable developers to better control their ethical and legal risks when using the datasets. (Research shows that 70%+ of licenses for popular datasets on GitHub and Hugging Face are "Unspecified". In addition, "the licenses that are attached to datasets uploaded to dataset sharing platforms are often inconsistent with the license ascribed by the original author of the dataset".)

- **Model license:** Legal documents that define the use and access of models through a contractual agreement between the model's provider (licensor) and the model user (licensee). Examples include open source licenses like MIT or Apache 2.0, custom licenses like OpenRAIL, and custom non commercial licenses like CC-BY-NC 4.0. More openness can enable developers to better control their ethical and legal risks when using the models or building on top of them (e.g. AI licensing categories from Open Core Ventures). To promote clarity in the usage of AI artifacts, it is important to distinguish between various components of the tech stack. For instance, between the model architecture, treated as software code, and the learned model parameters, which constitute data. While open-source licenses may suit code, they could prove inadequate for data like model parameters. This mismatch underscores the need for new licensing frameworks tailored to the unique characteristics of AI artifacts. Adhering to best practices, such as employing approved open-source licenses for code and open content licenses for data, can offer transparency and facilitate collaboration. This shift towards specialized licensing could benefit future AI system development.
- **Code license:** Legal documents that define the use and access of code through a contractual agreement between the code provider (licensor) and the code user (licensee). Like traditional software, code licenses define the rules to use, modify and redistribute the code. More openness via access to code licenses enable the developers to know the conditions and permissions before using the code.

## SAFEGUARDS

When it comes to safeguards, they encompass traditional software and platform safety best practices (e.g., usage policies and safety-by-design principles), technology safeguards adapted for AI systems (e.g., moderation, which can be applied throughout the stack, from training data to model outputs), and safeguards that seem unique to foundation models (e.g., refusals and safety fine-tuning).

## TYPE OF RELEASE

- **Gradual / Staged release** (how access is granted, and by whom): Whether the model is incrementally or gradually accessible to a broader set of actors. For example, third-party auditors or expert red-teamers can have earlier access to the model than the general public. Every stage of the release enables experts to scrutinize the model's usage, evaluate its societal impacts, and incorporate potential patches or heightened safety measures. This approach is being debated on whether a comprehensive assessment should be undertaken, unveiling the following, more advanced iteration.

- **Gated access or public access** (provided up to some conditions like registration vs. provided without any prerequisites): Gated access enables identified actors to download part of the models (e.g. weights and training code) or to provide vetted researchers, auditors, red-teamers, specific access via to fine-tuning API (e.g. "research API"). These privileged model access can be coupled with staged release.
- **Hosted inference endpoint** (whether or not in, addition to downloadable weights, an inference endpoint is offered): Access to inference endpoints in addition to downloadable weights enable the developer to benefit from optimized inference services (e.g. through REST API) while enabling them to fully customize the model they use. It enables the developers to fully utilize the open models.

## USER POLICY

- **Acceptable use policy** (document stipulating constraints and practices that a user must agree to for access to the model): As the development of those models is changing rapidly and the technology is general purpose, it is not easy to foresee how the technology will serve in some particular applications and use cases. It greatly helps downstream developers to explicitly document with sufficient details any acceptable use of the foundation models that are released.
- **Reporting & Redress mechanisms:** Feedback loops between users, broader impacted people, developers and model providers are critical to ensure right and redress for all the stakeholder and affected people. Mechanisms like incident databases are first steps toward documenting and reporting misuses and accidents with models.

## GUARDRAILS

- **Aligned weights:** Weights that have been further trained with preference data as a safety mitigation technique. In particular, aligned weights are weights of a base model that have been fine-tuned with a dataset that encapsulates content moderation examples or examples of how to answer or refuse to answer to various questions. More openness can enable a community of researchers to further test jailbreaking techniques and establish benchmarks to document the robustness of safety techniques.
- **Programmable guardrails:** Explicit handwritten rules and filters (as opposed to implicit learnt rules) that sit on top of a model for content moderation purposes. Examples include NeMo-Guardrails, Guidance, and generation-guided tools like Outlines. More openness can enable developers to use transparent and auditable content moderation techniques, as opposed to closed black box APIs to categorize hate speech or prevent AI systems from answering certain topics or prompts.
- **Safeguard models:** A safeguard model inspects output (or input) of base foundation models to further decide whether it is compliant with various rules and content moderation policies. Examples include llama-guard and fine-tuned lmsys-chat. More openness can enable red teamers, researchers, and the general public to evaluate the robustness of content moderation systems, identify potential vulnerabilities, and participate in advancing AI research and safety.

- **System prompts:** Pieces of text that are added to user prompts and that explicitly give content moderation rules to the models. Access to prompt libraries like PromptSource better help sharing these prompts. More openness can enable people to collectively determine which prompts are more efficient to guide the model and increase safety while not compromising on utility.

# APPENDIX II - Illustrations of some debated topics

## 1.1 Debated topics focused on openness

This section aims to present examples of key public conversations within the literature review.

TOPICS OF AGREEMENT

- Openness in AI is not a software-only issue. Access to software code is not sufficient to guarantee inspectability and reproducibility of the AI models.
- Open-source models alleviate various costs for pre-training and fine-tuning models.
- FMs should be more culturally diverse. FMs are too often primarily trained on English, and would benefit from being trained on other languages like Spanish, Arabic, etc. and local languages (Basque, etc.).
- Open Weights does not necessarily mean Open Source.

TOPICS OF CONTROVERSY

**AI Openness & developers' best practices:**

- What are minimum standards for responsible reusability?
- What are minimum standards for replicable evaluation?
- What are the minimum AI components to share/release to ensure reproducibility?
- How can developers best document artifacts so that it is reproducible?
- How can developers best document both released and unreleased AI components?
- How can developers approach building AI systems in an open way, such as by encompassing transparency, collaboration, and inclusivity in the process?
- What should be expected at the design stage from developers for responsible release and safety? Should expectations be different based on whether the purpose of development is commercial vs. noncommercial, research vs. deployment, or for specific societal purposes?

**AI Openness & distribution:**

- What kind of model-sharing standards should be promoted to support safe model distribution?
- Should open-source foundation models be released progressively, and if so, what kinds of risk analysis and prioritization should be done and for what types of models?
- How and for what should open-source FMs be assessed before putting them on the market?
- Are new licenses for open source AI/ open models needed or not?
- When is "open source capture" problematic and when is it not? How do we limit the problematic cases or features of "open source capture"
- Should the only licenses deemed open source be the ones recognized by the OSI?

**AI Openness & safety:**

- What characteristics of AI systems (e.g., compute, data, application domain) should be used to estimate the marginal risk directly from open sourcing a particular FM or particular elements of the AI stack?
- How can we deal with risks created by the ability to remove guardrails from an AI system when its model weights are widely available?

- Is an additional safety layer needed as part of an additional training step to align the model before releasing it open source?
- Can AI openness for some AI systems help regulators with developing rules that apply to non-open AI systems?
- What characteristics of AI openness would help regulators, journalists, and academics identify issues in AI systems more quickly? (E.g. What metrics should be documented in public reporting of red teaming results?) Can openness help auditing identifying vulnerabilities in proprietary AI?
- Would safety measures and methodologies be different considering the type of FMs (e.g. depending on the modalities, input, output type, etc.)?

**AI Openness & datasets:**

- What criteria should be met to consider datasets "open"?
- Is using "open datasets" for FM training a safety vector?
- How to make AI training and testing datasets openly available while safeguarding privacy and security?

Illustrations of points of controversy are further analyzed and detailed in Appendix II;

## 1.2 Emerging debates and roadblocks on AI openness

Should open-source FMs be released progressively?

The question arises whether conducting pre-market audits and post-deployment assessments by external experts be mandatory due to the rising capabilities of (open-source) foundation models. In August 2023, ahead of DEF CON 31, Accountable Tech, AI Now, and EPIC published "Zero Trust AI Governance", a framework that suggest that third-party audits for FMs should be implemented, as it offers greater efficiency than self-assessments (first-party) or contracted auditing (second-party). This approach promotes the idea that pre-market third-party conformity should be assessed by notified bodies and supported by accredited third-party auditors. Those audits would be based on full API and data access, and standardized assessment.

Should open-source FMs be assessed before putting them on the market?

According to some AI researchers' publications, the release of FMs should be compulsory on 'structured access', whereby limits are placed on a system's use, modification and reproduction for a downstream application. This approach is opposed to the simple 'beta test' of LLM directly to the public, like for ChatGPT in November 2022. A 'structured access' approach advocates rather for an initial stage release (or closed-source release) to enable a period of risk observation, with then the option to release the full model parameters once validated/audited (e.g. Auditing LLM: a three-layered approach). Validation can be a one time event or a continual validation that may take account of a changing environment. Checks may be proportional to the risk at a given time or in any given context. The question also arises whether access should be granted to vetted researchers.

Should the most advanced FMs be open sourced?

When building FMs, AI developers can decide to open-source their models, meaning to make it publicly available - under certain conditions depending on the license's permissions, conditions, and limitations - for anyone to access, study, modify and build AI systems and applications on top of it. Some AI scholars have concluded that publishing these FMs involves unacceptable trade-offs and believe they should not be open sourced (e.g., claiming that once open-sourced, malicious actors could disable and circumvent safeguards and introduce new dangerous capabilities via fine-tuning). These labs are either maintaining their models entirely private or employing a structured access approach to model sharing through their APIs, imposing user restrictions and measures like safety filters.

### How does open source apply to the various software components of FMs?

There is a lack of consensus among experts regarding the specific components of a model that must be shared for it to be classified as open-source. Should AI openness apply to the weights, floating point numbers in an LLM, or to training data, or to something else? There is common agreement that software code used for training, testing, and inference should be open-source. Same holds true for model weights (which cannot be strictly seen as software code) and model architecture. Debate also revolves around whether there could be restrictions on retraining artifacts, and redistributing them.

### Should the original dataset used to train the model be considered as part of the source code of the model?

Some developers consider that training would be the equivalent of compilation, whereas others explain that AI models can be retrained or fine-tuned without the original dataset. The question would then be which other components would be required to allow this model to be retrained or fine-tuned, without having access to the original dataset? Another important point is that the capability to aggregate large quantities of data should be available to everyone, not just a few big corporations that can pay on licensing deals or that have already accumulated large volumes of data through their pre-existing services.

### Can AI openness help regulate proprietary AI?

Regulators need to close the gap in AI expertise and have an in-depth understanding of how to test AI systems, evaluate their impact on society and to govern them effectively. There is a strong need to equip supervising authorities adequately and efficiently to allow them to properly monitor, regulate and enforce AI. Supervisory bodies in charge of monitoring and enforcing AI regulations should have access to shared AI tools for regulators and other democratic stakeholders based on open-source models. Indeed, a "white box" approach could enable those players to lead by example by using tools that can be audited without relying on black box providers subject to such assessments.

### What are minimum standards for reproducibility of FM research?

Some have argued that ML has entered a reproducibility crisis. One of the main claims is that it is difficult to reproduce experiments and evaluations for independent researchers and third parties. For instance, a simple example from research can inform on the fact that rescoring all models from HELM (a widely respected living benchmark) regarding toxicity metrics, with recent versions of the HELM API lead to non-consistent rankings of widely used FM overtime. As this example illustrates, understanding downstream consequences of opaque or undocumented changes in the evaluation pipelines or in the training pipelines puts barriers to entry for researchers trying to reproduce results or to robustify them overtime. Other debates have emerged regarding the reproducibility of evaluation results for FMs, and how to enhance factuality (detect factual errors) in generative AI systems. In the case of LLMs, as in-context abilities have emerged, prompting engineered techniques have been developed to increase the performance of the models up to a point where we are now unable to meaningfully compare the evaluation results until custom prompts are revealed and documented. As there is no official "prompt" or way to evaluate them with LLMs, other researchers have argued that releasing evaluation prompts was optional. In addition, lack of reproducibility also emerges from some closed providers' business practices of deprecate and removing access to older models

## What kind of safety and security by design requirements should be promoted before releasing a model openly?

Training with human preference datasets has become one of the key design principles both to better align the FMs on users preferences and also as part of an embedded safety layer where alignment helps reduce models toxicity and potential hate speech and can reduce biases. While some argue there is no tradeoff between model's safety and model's utility when training with user preference data, other developers note that alignment techniques make the model collapse on less diverse outputs and a more narrow vision of the world. In addition, safety researchers have proved that if the weights of a model are released, safeguarding it with current alignment methods is worthless as the alignment can be easily broken with very cheap additional training by a malicious actor. As a result, it has been debated whether or not base models should be trained to comply with base content moderation rules and thus deployed with embedded safety layer(s). Some of the opponents to this statement argue that there is no "one true correct alignment" and that giving access to a diversity of foundation models is critical to ensure all cultures and views of the world are represented. Others argue that the choice of safety layers depend on the kind of harms one wants to mitigate: model's alignment would protect against accidental harms not intentional ones.

## APPENDIX III - Overlay with the Model Openness Framework

The Model Openness Framework (MOF), published in March 2024 by the Linux Foundation, is a ranked classification system that rates ML models based on their completeness and openness. The MOF requires specific components of the model development lifecycle to be included and released under appropriate open licenses. There are 17 items in the MOF. More precisely, 15 components that come from the model development lifecycle and those artifacts are associated with different classes of the MOF. In addition to the MOF components, the framework requires the inclusion of a master inventory file, MOF.JSON, which includes information on the components included and their corresponding open licenses.

These 17 items will be described as follows:

- MOF1. Datasets
- MOF2. Data Preprocessing Code
- MOF3. Model Architecture
- MOF4. Model Parameters
- MOF5. Model Metadata
- MOF6. Training, Validation and Testing Code
- MOF7. Inference Code
- MOF8. Evaluation Code
- MOF9. Evaluation Data
- MOF10. Evaluation Results
- MOF11. Supporting Libraries and Tools
- MOF12. Technical Report
- MOF13. Model Card
- MOF14. Data Card
- MOF15. Research Paper
- MOF16. Sample Model Outputs
- MOF17. Model Openness Framework Configuration File

The below figure compares the 17 items of the MOF with the framework developed in this paper, by overlaying the two, showing how the Model Openness Framework can be considered a normative layer applied to the descriptive framework outlined in this paper.

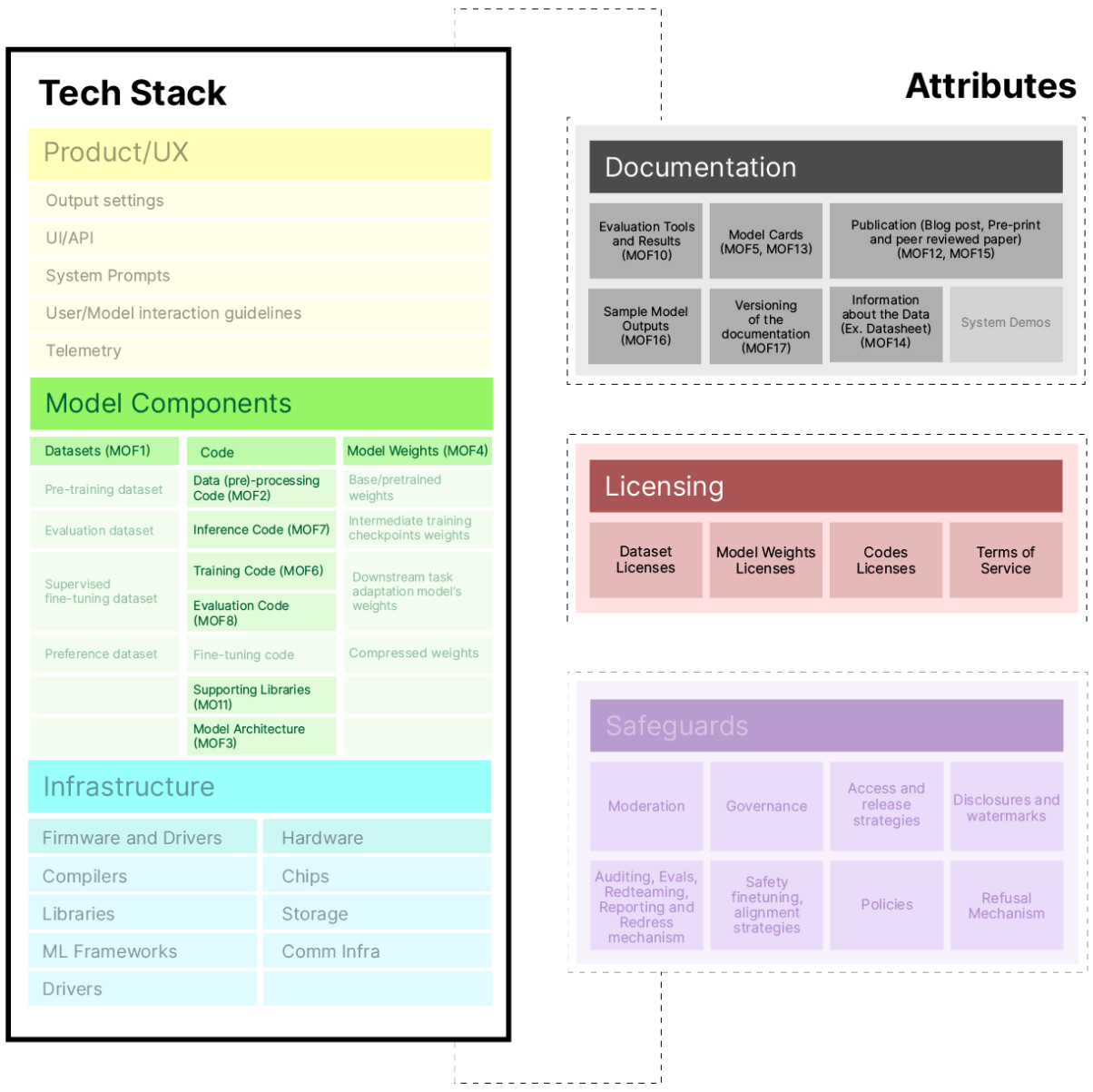

**Fig 12: Comparison of Dimensions of AI Openness Framework with the MOF**

Another normative approach developed by the Open Source Initiative, who participated in the Columbia Convening, posits that an AI system is "open source" when it is "made available under terms that grant the freedoms to **use** the system for any purpose and without having to ask permission, **study** how the system works and inspect its components, **modify** the system for any purpose, including to change its output, and **share** the system for others to use with or without modifications, for any purpose.[14]" reanchoring the four freedoms of free program by Richard Stallman.

Building on the Model Openness Framework cited above, the normative definition proposed by OSI takes 11 of these core components and assorts them with licensing requirements in order to meet the criteria for "open source AI".

[14]The Open Source AI Definition version 0.0.8, OSI, April 2024